\documentclass[twocolumn,secnumarabic,amssymb, nobibnotes,superscriptaddress, aps, prb]{revtex4-1}
\usepackage{amsmath,amssymb,graphicx,color,bm,soul}
\usepackage{ulem}
\begin{document}

\title{Vortex-vortex control in exciton-polariton condensates}

\author{Xuekai Ma}
\affiliation{Department of Physics and Center for Optoelectronics and Photonics Paderborn (CeOPP), Universit\"{a}t Paderborn, Warburger Strasse 100, 33098 Paderborn, Germany}

\author{Stefan Schumacher}
\affiliation{Department of Physics and Center for Optoelectronics and Photonics Paderborn (CeOPP), Universit\"{a}t Paderborn, Warburger Strasse 100, 33098 Paderborn, Germany}

\affiliation{College of Optical Sciences, University of Arizona, Tucson, AZ 85721, USA}

\begin{abstract}
Vortices are widely studied in fields ranging from nonlinear optics to magnetic systems and superconductors. A vortex carries a binary information corresponding to its topological charge, `plus' or `minus', that can be used for information storage and processing. In spatially extended optical and condensed many-particle systems, achieving full control over vortex formation and its charge is particularly difficult and is not easily extended to systems of multiple vortices. Here we demonstrate the optical creation of multiplets of phase-locked vortices in polariton condensates using off-resonant excitation with ring-shaped pump beams. We find that the vorticity of one vortex can be controlled solely using the phase-locking with other nearby vortices. Using this mechanism, we demonstrate how an existing vortex with a specific topological charge can be inverted to the oppositely charged state, and how the charge state of one reference vortex can be copied to a neighboring vortex. This way we can optically encode any set of binary information onto a chain of vortices. We further show that this information can be modified later by using the possibility to address and manipulate each vortex in the chain individually.
\end{abstract}


\maketitle

\section{introduction}
Vortices are widely studied in different areas of physics, including superconductors~\cite{Bardeen-pr-1965,Miyahara-apl-1985,Blatter-rmp-1994,Harada-science-1996,Roditchev-nphy-2015}, magnetic system~\cite{Shinjo-science-2000,Waeyenberge-nature-2006,Pribiag-nphy-2007,Cowburn-2007-nmat,Kim-2008-apl}, atomic condensates~\cite{Dum-prl-1998,Matthews-prl-1999,Madison-prl-2000,Isoshima-pra-2000,Weiler-nature-2008}, and nonlinear optics~\cite{Swartzlander-prl-1992,Alexander-prl-2004,Ferrando-oe-2004,Briedis-oe-2005}. A vortex consists of a core, where the phase is singular and the density of particles reaches its minimum, and a circular flow around the core, with the phase winding being an \textcolor{black}{integer multiple of $2\pi$~\cite{Pitaevskii-book-2003}}. The sign of the vortex charge is positive or negative representing different rotation directions. These differently charged states of a vortex can be used to store and process information in a binary fashion. The control of vortices was previously achieved in superconductors~\cite{Miyahara-apl-1985} and magnetic materials~\cite{Cowburn-2007-nmat,Kim-2008-apl}, encouraging the development of a vortex-based random access memory. In the past decade, remarkable progress was also made in the study of vortices in polariton fluids~\cite{Marchetti-prl-2010,Lagoudakis-prl-2011,Pigeon-prb-2011,Tosi-ncomm-2012,Dagvadorj-prx-2015,Boulier-sr-2015} and condensates~\cite{Rubo-prl-2007,Lagoudakis-science-2009,Manni-ncomm-2012,Keeling-prl-2008,Lagoudakis-nphys-2008,Sanvitto-npho-2011,
Roumpos-NatPhys-2011,Manni-prb-2013,Dall-prl-2014,Sigurdsson-prb-2014,Ma-prb-2016,Ma2016symbiotic,Liew-prb-2015}. Polaritons are quasi-particles in semiconductor microcavities that form as composites of cavity photons and quantum-well excitons. Through their photonic component they posses long-lived coherence and can be optically excited and through their excitonic component they interact, giving rise to prominent nonlinearities. For incoherent excitation, spontaneous formation of macroscopic quantum coherence in a low-energy state of polaritons was observed \cite{Deng-science-2002,Kasprzak-nature-2006,Balili-science-2007}, commonly referred to as Bose-Einstein condensation of polaritons. Unlike atomic condensation, polariton condensation can be achieved not only at cryogenic temperatures but up to room temperature~\cite{Lidzey-prl-1999,Zamfirescu-prb-2002,Malpuech-apl-2002}, rendering polariton condensates promising for polaritonic devices for all-optical information processing.

\begin{figure}
\includegraphics[width=.9\columnwidth]{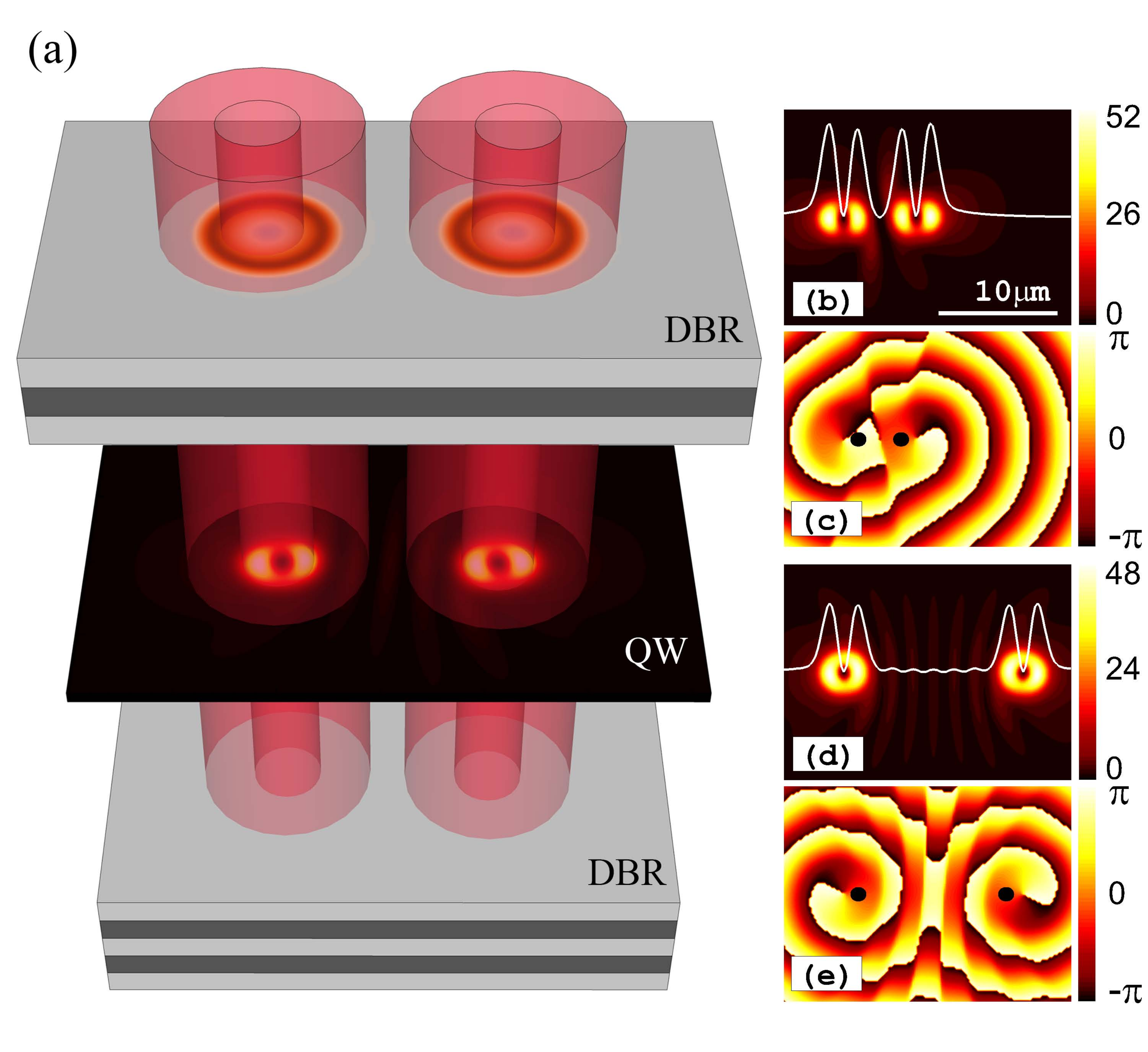}
\caption{(Color online) (a) Sketch of a semiconductor microcavity with excitation by two off-resonant ring-shaped pump beams. The microcavity system consists of a semiconductor quantum well (QW) sandwiched between two distributed Bragg reflectors (DBRs). Sample distributions of (b,d) density (in $\mathrm{\mu m^{-2}}$) and (c,e) phase of two phase-locked vortices at distance of (b,c) $d=6\,\mathrm{\mu m}$ and (d,e) $d=15\,\mathrm{\mu m}$ are shown. White lines in (b,d) are profiles of the density along the line connecting the centers of the two vortices. Phase differences of the vortices in (c) and (e) are analyzed at the positions marked by the black points as discussed in the text.}\label{f1}
\end{figure}

The vortices in polariton condensates can be grouped into two types, cyclone-like vortices and crater-like vortices, respectively. Cyclone-like vortices are generated for spatially homogeneous~\cite{Liew-prb-2015,Ma2016solitons} or broader~\cite{Lagoudakis-science-2009,Manni-ncomm-2012,Keeling-prl-2008,Lagoudakis-nphys-2008} pump sources due to fluctuations during condensation. These vortices form randomly from initially present phase defects, and they always appear as vortex-antivortex pairs to preserve the system's total angular momentum. Crater-like vortices can be created using a spatially inhomogeneous pumping source, e.g. a ring-shaped optical pump~\cite{Sigurdsson-prb-2014,Ma-prb-2016} or a pump with circular intensity profile~\cite{Dall-prl-2014}. These profiles have been demonstrated to efficiently generate polariton consensates and can be generated using a spatial light modulator~\cite{Schmutzler2015,Niemitz2016}. The pump profiles generate a confining potential, which forces the condensate to assume a state with ring-shaped density profile with a phase dislocation in the center. These vortices are not very sensitive to noise and their location is determined by the pump profile. The winding number of these vortices depends on the specific pump profile used. For sufficiently small pump profiles only vortices with charge 1 form which are very stable. The sign of their topological charge is either `+' or `-' depending on initial noise. To break the radial system symmetry and to control the rotation direction of vortices, different approaches were used in previous studies including chiral polaritonic lenses~\cite{Dall-prl-2014}, external potentials~\cite{Sigurdsson-prb-2014}, and elliptical pump sources~\cite{Ma-prb-2016}. However, these approaches are not easily extended to flexibly control multiple coupled vortices in an array.

\begin{figure} 
\includegraphics[width=1\columnwidth]{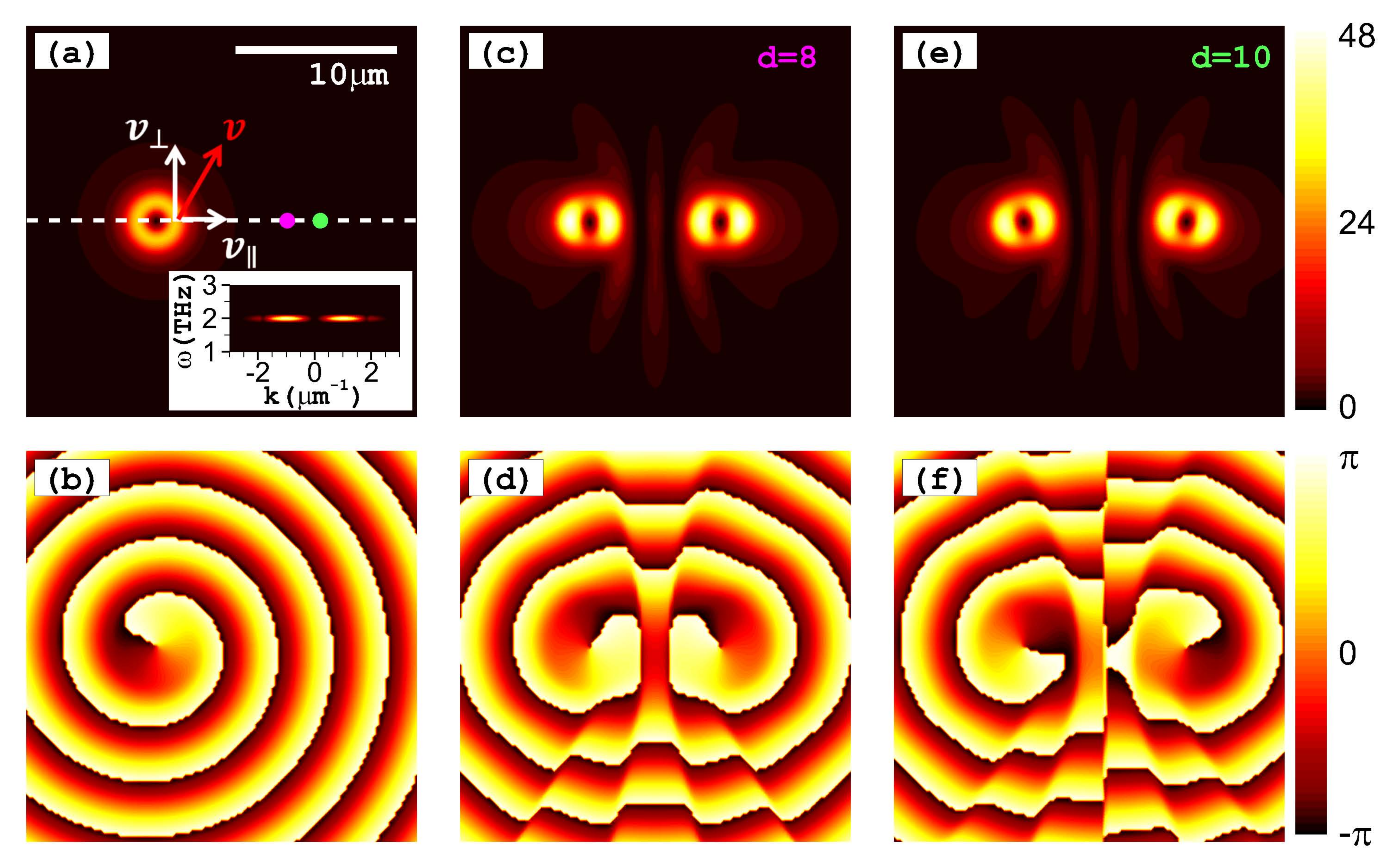}
\caption{(Color online) Distributions of density (upper row, in $\mathrm{\mu m^{-2}}$) and corresponding phase (lower row) of vortices. (a) and (b) show a single vortex. The inset in (a) shows the frequency and momentum distribution of the density taken along the dashed white line. The arrows in (a) represent the velocity ($v$) of polaritons and its radial ($v_\parallel$) and tangential ($v_\perp$) components. The two points in (a) indicate the centers of the additional ring-shaped pumps used in (c) for distance $d=8\,\mathrm{\mu m}$ (indicated in pink) and in (e) for  distance $d=10\,\mathrm{\mu m}$ (indicated in green). Density and phase of two phase-locked vortices are shown for distances $d=8\,\mathrm{\mu m}$ in (c) and (d) and $d=10\,\mathrm{\mu m}$ in (e) and (f), respectively.}\label{f2}
\end{figure}

\begin{figure} 
\includegraphics[width=1\columnwidth]{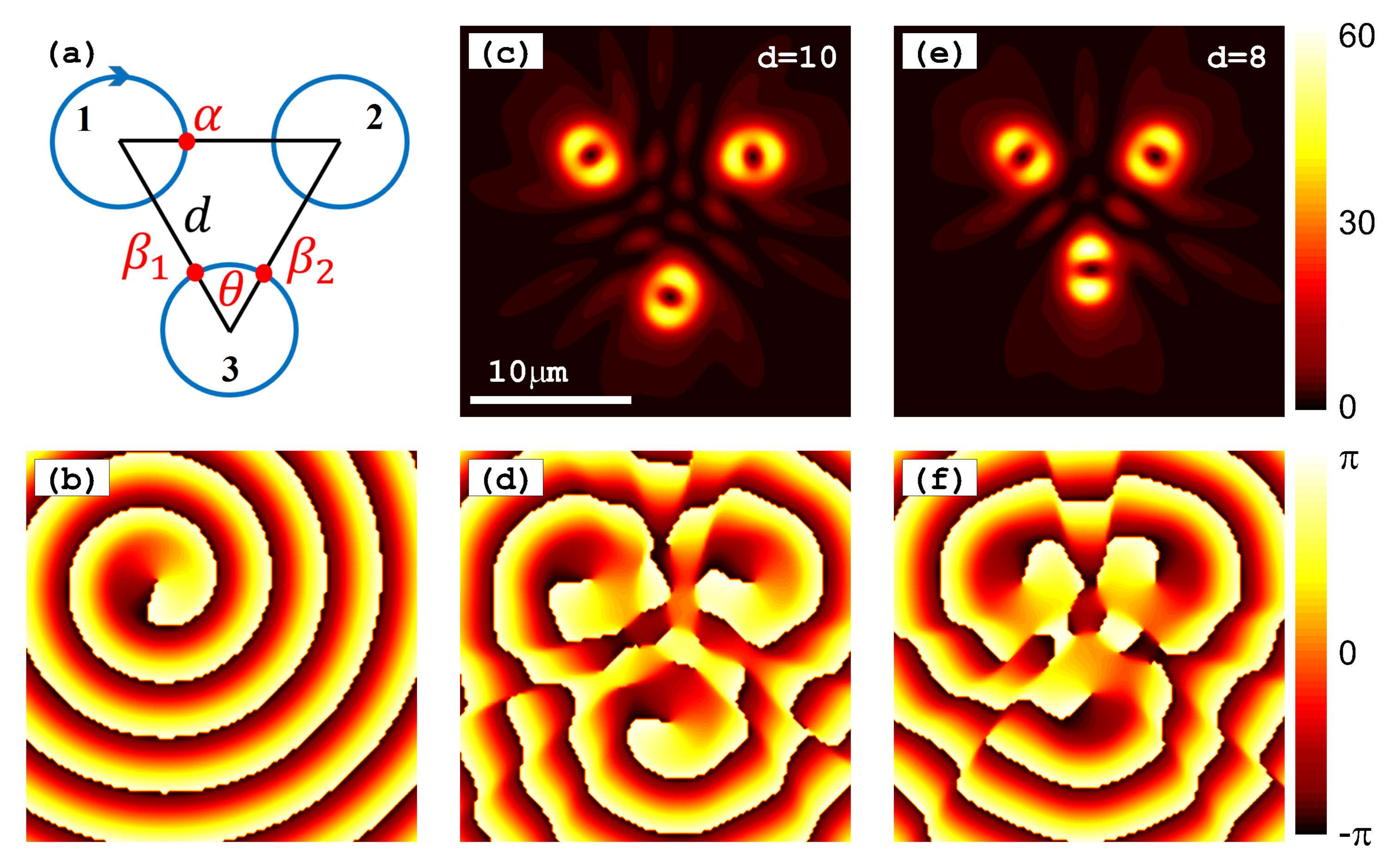}
\caption{(Color online) (a) Sketch of three phase-locked vortices arranged on a regular triangle. $\alpha$, $\beta_1$, and $\beta_2$ are the phase values at the points as indicated by the red points. $\theta=\pi/3$ is the internal angle of the regular triangle. (b) Phase distribution of a single vortex. Distributions of (c) density (in $\mathrm{\mu m^{-2}}$) and (d) phase of three phase-locked vortices with distance $d=10\,\mathrm{\mu m}$. Distributions of (e) density and (f) phase of three phase-locked vortices with $d=8\,\mathrm{\mu m}$.}\label{f3}
\end{figure}

Previously it was shown that when two spatially localized polariton condensates are sufficiently close to each other, they form a phase-locked state with a phase difference $0$ or $\pi$ between the condensate wave functions depending on their distance and outflow velocity of the condensates~\cite{Ohadi-prx-2016}. This for example can be exploited in polariton simulators~\cite{Berloff-aXiv-2016}. This phase locking of spatially separated condensates similarly applies to vortices. As illustrated in Fig.~\ref{f1}, each vortex is generated inside the ring-shaped profile of an off-resonant pump. The phase difference between neighboring vortices is then either $0$ or $\pi$ depending on their distance. We show that this phase locking provides a means to also control the topological charge of one vortex through the charge of the other. We find that two coupled vortices always carry opposite topological charge (cf. Fig.~\ref{f2}). For three coupled vortices in a triangular arrangement with a specific distance between them (here 10 microns), the vortices have the same topological charge (cf. Fig.~\ref{f3}). Utilizing these two observations, below we show as a potential application that any set of binary information can be encoded onto a chain of vortices. We further show that this information can be modified subsequently using the possibility to manipulate each vortex bit individually, using only off-resonant continuous wave optical excitation.

\section{theoretical model and phase-locked vortices}
To describe the dynamics of an exciton-polariton condensate that spontaneously forms in the vicinity of the bottom of the lower polariton branch, we use a mean-field theory. The dynamics of the condensate is coupled to an exciton reservoir created by an incoherent pump source. The coupled equations of motion for polariton field $\Psi(\mathbf{r},t)$ and reservoir density $n(\mathbf{r},t)$ read~\cite{Wouters-prl-2007}:
\begin{equation}\label{e1}
\begin{aligned}
i\hbar\frac{\partial\Psi(\mathbf{r},t)}{\partial t}=&\left[-\frac{\hbar^2}{2m}\nabla_\bot^2-i\hbar\frac{\gamma_c}{2}+g_c|\Psi(\mathbf{r},t)|^2 \right.\\
&+\left.\left(g_r+i\hbar\frac{R}{2}\right)n(\mathbf{r},t)\right]\Psi(\mathbf{r},t),
\end{aligned}
\end{equation}
\begin{equation}\label{e2}
\frac{\partial n(\mathbf{r},t)}{\partial t}=\left[-\gamma_r-R|\Psi(\mathbf{r},t)|^2\right]n(\mathbf{r},t)+P(\mathbf{r},t).
\end{equation}
Here $m=10^{-4}m_e$ (with $m_e$ being the free electron mass) is the effective mass of polaritons on the lower branch. The polaritons and excitons decay with $\gamma_c=0.33\,\mathrm{ps^{-1}}$ and $\gamma_r=1.5\gamma_c$, respectively. The strength of polariton-polariton and polariton-exciton interaction is given by $g_c=6\times10^{-3}\,\mathrm{meV\mu m^2}$ and $g_r=2g_c$, respectively. $R=0.01\,\mathrm{ps^{-1}\mu m^2}$ determines the replenishment of the condensate from the reservoir. These parameters correspond to the experimental data reported in \cite{Roumpos-NatPhys-2011}. The incoherent reservoir $n(\mathbf{r},t)$ is created by an off-resonant pump source $P(\mathbf{r},t)$ spectrally well above the band gap. We consider a ring-shaped continuous-wave (CW) pump with the following spatial profile:
\begin{equation}\label{e3}
P(r)=P_0\left[1-e^{{-\left(\frac{r}{w}\right)^2}}\right]e^{-\left(\frac{r}{w}\right)^{10}},
\end{equation}
with $r=\sqrt{x^2+y^2}$, $P_0=150\,\mathrm{ps^{-1}\mu m^{-2}}$, and $w=3\,\mathrm{\mu m}$. In our previous work we found that with $w=3\,\mathrm{\mu m}$ only persistent vortices with topological charge $M=\pm1$ are generated inside the ring~\cite{Ma-prb-2016}.

When two ring-shaped pumps with the same spatial profile are present as illustrated in Fig.~\ref{f1}(a), two phase-locked vortices are formed. Besides the two reservoir-induced 2D parabolic potentials confining each vortex, there is an additional quasi-1D potential between the two rings. The width of this potential depends on the distance between the two pumps. It was previously demonstrated that when the potential is sufficiently narrow, the two fundamental condensate modes (without any vortex being present) have equal phase~\cite{Ohadi-prx-2016,Berloff-aXiv-2016}. With increasing distance between the confined condensates, the  wavefunction shows an increasing number of mini maxima as in Fig.~\ref{f1}(d). As a consequence of the overall symmetry of the wavefunction of the coupled system, for an even number of mini maxima the confined main parts of the condensates have a phase difference of $\pi$, for an odd number of mini maxima the phase difference is $0$. We note that when the two pumps are far away from each other, polaritons simultaneously occupy several levels in the potential energy landscape leading to complex temporal oscillations in real space~\cite{Tosi-np-2012}. We find that this phase locking similarly applies to two vortices and also depends on the distance between the two traps. The distance $d$ is the distance between vortex cores in the centers of the two ring-shaped pumps. When the distance is small, for instance $d=6\,\mathrm{\mu m}$ as shown in Figs.~\ref{f1}(b) and (c), there is no mini maximum (corresponding to an even number) between the main peaks of the two vortices. The phase difference is $\pi$ (compare the phases as marked by the black points in Fig.~\ref{f1}(c)). In this case, the two vortices strongly interact with each other leading to a more dipole-like profile.

Increasing the distance to $d=15\,\mathrm{\mu m}$, the interaction between the two vortices becomes weaker and five mini maxima (odd number) are formed as shown in Fig.~\ref{f1}(d). In this case, the phase of the two vortices is equal [marked by the black points in Fig.~\ref{f1}(e)]. This shows that two nearby vortices can form a phase-locked vortex pair for which the phase difference depends on the distance between vortices. Here we define the phase difference between two vortices as $0$ or $\pi$, respectively, depending on the difference in phase at the main density maxima on the line between the centers of the two vortices as introduced in Fig.~\ref{f1}(c) and (e). It is important to note that even if the phases of two vortices are locked together as discussed above, in general if generated simultaneously the two vortices can still have any combination of different or equal topological charges, depending on noise during the formation. Deterministically controlling the topological charges of phase locked vortices is the main aim of the present work.


\section{vortex-vortex control}
First we demonstrate the controlled creation of phase-locked vortices with opposite topological charges. We define that a vortex rotating counter-clockwise has positive charge, $M=+1$, and a vortex rotating clockwise has negative charge, $M=-1$. When for a single ring-shaped pump, a vortex with $M=+1$ is formed as in Figs.~\ref{f2}(a) and (b), the  polaritons in the condensate propagate with a radial velocity ${v_\parallel}$ and a tangential rotation-induced velocity ${v_\perp}$ which is related to the vorticity of the condensate as indicated in Fig.~\ref{f2}(a). The two velocity components are given by
\begin{equation}
v_\parallel=\frac{\hbar k}{m} \qquad \text{and} \qquad \,
v_\perp=\omega r'
\end{equation}
with the wavenumber $k$ of polaritons, the angular frequency $\omega$ of the vortex and its radius $r'$. The values of these quantities can be extracted from Fig.~\ref{f2}(a) and the inset. For the maximum in density, we obtain $\omega\simeq2.0\,\mathrm{THz}$, $k\simeq1.0\,\mathrm{\mu m^{-1}}$, and $r'=1.15\,\mathrm{\mu m}$. This leads to $v_\parallel\simeq1.16\,\mathrm{\mu m/ps}$ and $v_\perp\simeq2.3\,\mathrm{\mu m/ps}$ which are of similar magnitude, leading to an outgoing polariton flow with total velocity $v=\sqrt{v_\parallel^2+v_\perp^2}$ along the direction indicated by the red arrow in Fig.~\ref{f2}(a). Thus the polaritons flowing from the condensate break the radial symmetry of the system as seen by a nearby vortex formed at a later time. Due to the direction of the flow for a first vortex with counter-clockwise rotation, $M=+1$, the second vortex always rotates clockwise with a topological charge $M=-1$ for both the $0$-state as in Figs.~\ref{f2}(c) and (d) and the $\pi$-state as in Figs.~\ref{f2}(e) and (f). In conclusion, the second  vortex (formed at a later time) always has topological charge opposite to the topological charge of the first vortex.

Now let us consider the controlled generation of two phase-locked vortices with the same topological charge. To this end we consider three ring-shaped pumps arranged in a regular triangular geometry as shown in Fig.~\ref{f3}(a). To analyze the relation of the three topological charges generated, we assume the first vortex to be negatively charged with $M\equiv-1$ as indicated by the arrow in Fig.~\ref{f3}(a). For phase-locking in the $\pi$-state, if the second vortex has the same topological charge with $M=-1$, the phases $\beta_1$ and $\beta_2$ in the third vortex can be given relative to the phase $\alpha$ of the first vortex:
\begin{equation}
\beta_1=\alpha+\frac{\pi}{3}+\pi   \quad \text{and} \quad  \beta_2=\alpha+\pi-\frac{\pi}{3}+\pi\,.
\end{equation}
So the difference of the phases $\beta_1$ and $\beta_2$ is
\begin{equation}\label{e8}
\beta_1-\beta_2=-\frac{\pi}{3},
\end{equation}
that is, $|\beta_1-\beta_2|=\theta\equiv\pi/3$. For $\beta_1-\beta_2<0$ the third vortex also rotates clockwise with the topological charge $M=-1$, so that the three vortices have the same topological charge as in Figs.~\ref{f3}(c) and \ref{f3}(d). If the second vortex had topological charge $M=+1$, the phase difference in the third vortex would be $|\beta_1-\beta_2|=\pm\pi\neq\theta$. This is not compatible with the geometrical arrangement of the vortices such that the second vortex can not have opposite topological charge to the first one. For phase-locking in the $0$-state, if the second vortex has topological charge $M=-1$, the phase difference in the third vortex is $|\beta_1-\beta_2|=2\pi/3\neq\theta$. While if the second vortex has topological charge $M=+1$, the phase difference in the third vortex is $|\beta_1-\beta_2|=0\neq\theta$. In this case there is no pre-determined phase relation for the three vortices with $0$-state phase locking. In conclusion, we find that for the three phase-locked vortices arranged on a regular triangle, for $\pi$ phase locking the vortices always have the same  topological charge, cf. Figs.~\ref{f3}(c) and (d). As a consequence, when a single vortex is formed first, say with the topological charge $M=-1$ as in Fig.~\ref{f3}(b), the two vortices formed simultaneously at a later time also have topological charge $M=-1$. In other words, the topological charge of the first vortex is copied onto the other two vortices such that their phases are synchronized. For vortices with $0$ phase locking, however, the three topological charges do not follow from the analytical considerations above and do not synchronize as shown in Fig.~\ref{f3}(e) and (f).

\begin{figure} 
\includegraphics[width=1\columnwidth]{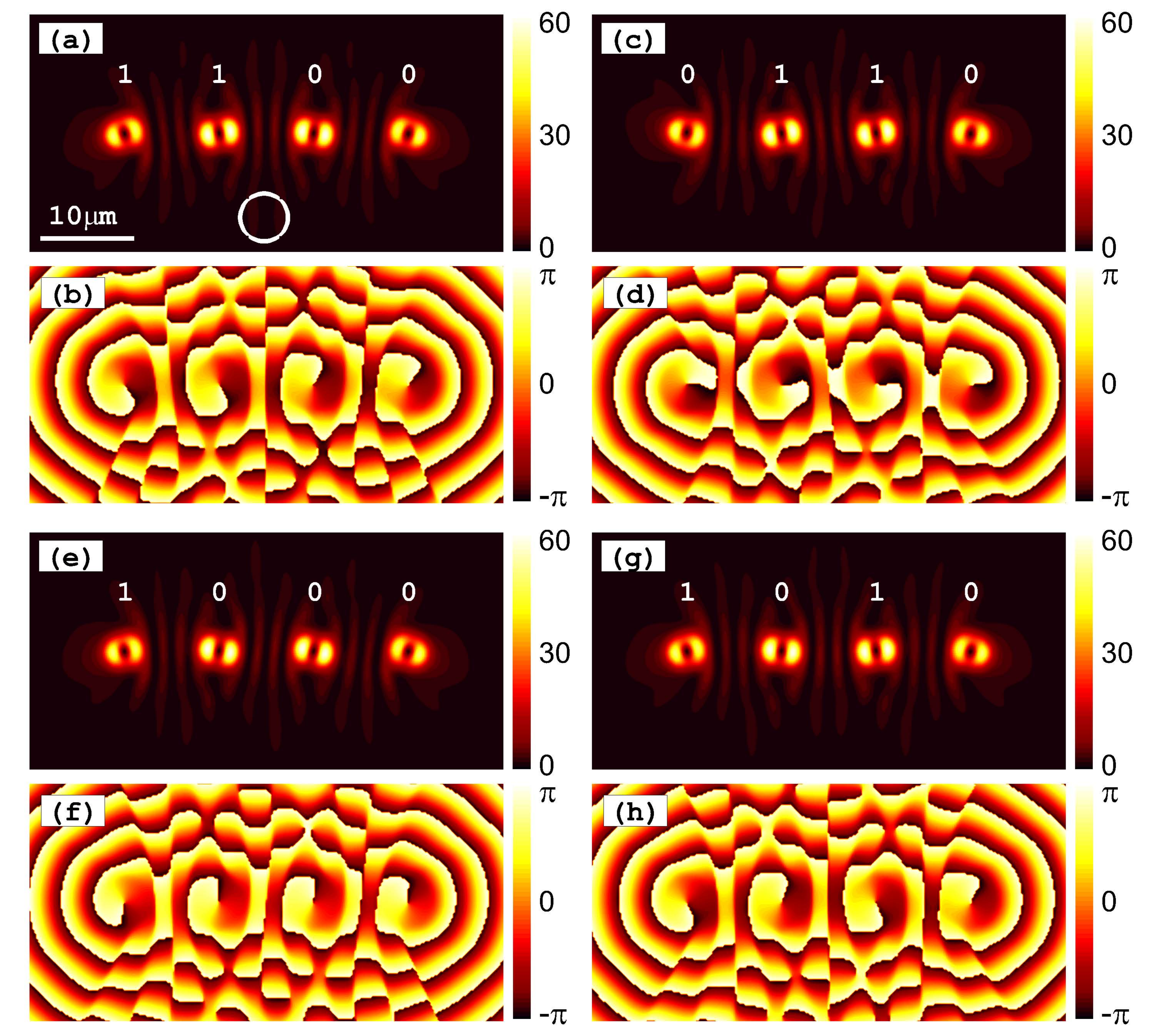}
\caption{(Color online) A vortex chain containing four vortices phase-locked in the $\pi$-state with distance $d=10\,\mathrm{\mu m}$. Distribution of (a,c,e,g) density (in $\mathrm{\mu m^{-2}}$) and (b,d,f,h) corresponding phase. (a) and (b) show the vortices in the `1 1 0 0' state. (c) and (d) show the `0 1 1 0' state. (e) and (f) show the `1 0 0 0' state. (g) and (h) shwo the `1 0 1 0' state. The white circle in (a) indicates the position of an additional control pump used to switch between different vortex configurations.}\label{f4}
\end{figure}

From the discussion above it follows that with one predefined vortex an oppositely charged vortex can be generated for both $0$-state and $\pi$-state phase locking. Generation of a second vortex with the same charge as the first vortex can be achieved using an additional control vortex (which can be switched off after generation is completed without perturbing the other two vortices). This can only be done for $\pi$-state phase locking. Therefore, for $\pi$-state phase-locked vortices we are able to generate a second vortex with pre-determined charge, + or -. In Fig.~\ref{f4}(a) we generated a chain of four vortices (four memory-bits) in a pre-determined binary sequence of `1 1 0 0'. Here we define that topological charge $+1$ corresponds to the `1' state and topological charge $-1$ corresponds to the `0' state of each bit. These four memory bits are composed of four vortices with $\pi$-state phase-locking for $d=10\,\mathrm{\mu m}$. Below we show that after generation of the chain of vortices each bit can independently be switched to its opposite charge such that the information stored can be changed without re-generating the entire chain. As the vortices are quite close to each other with a strong correlation between them, the best way to manipulate one of them is to use the neighboring vortices (or memory bits).

\section{manipulation of a vortex chain memory}
In Fig.~\ref{f4}(a) and (b) the first two vortices have the same topological charge such that both bits are in the `1' state. To switch the first bit to its opposite state, we can use the method introduced in Fig.~\ref{f2}. Thus we switch off the first vortex  (for some tens of picoseconds or longer) until it disappears. Then we switch the source back on again. Due to the influence of the second vortex, the newly created first vortex (or bit) now is in the `0' state as shown in Figs.~\ref{f4}(c) and (d). For a vortex that is sandwiched between two vortices with different charges, for example the third vortex in Fig.~\ref{f4}(a), the vorticity can only be switched using a third vortex as introduced in Fig.~\ref{f3}. To switch the third vortex in Fig.~\ref{f4}(a) to the `1' state, i.e., the same state as the second vortex we first switch off the source for vortex three and then switch it back on and simultaneously introduce an additional ring-shaped pump at the position marked by the white ring in Fig.~\ref{f4}(a) with a distance $d=10\,\mathrm{\mu m}$ to the second and the third vortex. After formation of the vortices on the chain, the additional pump can be switched off. In this scenario, the information stored in the second vortex is copied onto the third vortex as shown in Figs.~\ref{f4}(c) and (d). We note that the presence of the forth vortex does not disturb this approach such that the `1 1 0 0' state [Figs.~\ref{f4}(a) and (b)] can be switched to the `0 1 1 0' state [Figs.~\ref{f4}(c) and (d)].

The second vortex in Fig.~\ref{f4}(a) is sandwiched between two vortices with opposite charges. Analogously, if switching off the second vortex for a period of time and then switching it on again together with the additional control, the information of the third vortex is copied onto the second vortex resulting in the `1 0 0 0' state [Figs.~\ref{f4}(e) and (f)]. For the `1 0 0 0' state the third vortex is sandwiched between two vortices with equal charges. As introduced in Fig.~\ref{f2} the charge of the third vortex can be inverted by switching it off and back on again, resulting in the `1 0 1 0' state [Figs.~\ref{f4}(g) and (h)]. To switch the third vortex back to the `1 0 0 0' state, an additional control pump is needed to copy the topological charge from the second or the forth vortex. In our work, the total time for switching a vortex charge between different states is on the order of several hundreds of picoseconds, which is related to the decay time and the formation time of a vortex. We note, however, that this timescale depends on system parameters and will in particular decrease with increasing pumping power. It is worthwhile noting that the same scheme demonstrated above can be extended to vortex memory bit chains of arbitrary length. Although crater-like vortices have not been discussed or observed for electrically pumped condensates~\cite{Tsintzos-nature-2008,Schneider-nature-2013,Bhattacharya-prl-2014}, we believe that electrically controlled vortex memory chains may be an interesting subject for future studies.

\section{conclusion}
In the present work we investigated phase locking of vortices in polariton condensates generated with incoherent ring-shaped pump beams. We show that two simultaneously formed neighboring vortices can have either the same or opposite topological charges. A predefined vortex with certain vorticity, however, breaks the radial system symmetry such that a second vortex formed in its vicinity at a later time always has the opposite topological charge. Three vortices locked in the $\pi$ phase state in a regular triangular arrangement, all have the same topological charge as a consequence of phase coupling and geometrical arrangement. As a possible application, we demonstrate that together these properties can be utilized to realize an optically controlled memory based on a chain of vortices. These results open up exciting avenues for future research into phase locked vortices and their application to all-optical information processing. Also vortex-vortex control in other geometries may be of interest for future studies, e.g., in a hexagon configuration.

\section*{acknowledgments}
This work was supported by the Deutsche Forschungsgemeinschaft (DFG) through the collaborative research center TRR 142 and the Heisenberg program. We further acknowledge computing time on the clusters of the Paderborn Center for Parallel Computing, $\mathrm{PC^2}$.

\end{document}